\documentclass[runningheads]{llncs}
\usepackage[T1]{fontenc}
\usepackage{graphicx}
\usepackage{amsmath,amssymb,mathrsfs, eucal}
\usepackage{algorithm2e}
\usepackage{multirow}
\usepackage{ctable} 
\usepackage{microtype}
\RestyleAlgo{ruled}
\usepackage{amsfonts}

\newcommand{\Z}{\mathbb{Z}}

\newcommand{\F}{\mathbb{F}}

\newcommand{\ord}{{\text {\rm ord}}}

\newcommand{\calA}{\mathcal{A}}

\newcommand{\calE}{\mathcal{E}}

\newcommand{\abs}[1]{\left\vert#1\right\vert}

\newcommand{\al}{\alpha}

\newcommand{\ve}{\varepsilon}

\newcommand{\ZZ}{\mathbb{Z}}
\newcommand{\sk}{\mathsf{sk}}

\renewcommand{\pmod}[1]{\  \,  \left(  \operatorname{mod} \,  #1 \right)}

\def\vp{\varphi}
\def\al{\alpha}
\def\vt{\vartheta}

\begin{document}
\title{Decomposition of RSA modulus applying even order elliptic curves}
%
%
\author{Jacek Pomykała\orcidID{0000-0002-6480-5751} \and
Mariusz Jurkiewicz\inst{1}\orcidID{0000-0002-6314-4381}}
\authorrunning{J. Pomykała et al.}
%
\institute{Military University of Technology, Kaliski 9, Warsaw, Poland 
\email{\{jacek.pomykala,mariusz.jurkiewicz\}@wat.edu.pl}}
\maketitle              
\begin{abstract}
An efficient integer factorization algorithm would reduce the security of all variants of the RSA cryptographic scheme to zero. Despite the passage of years, no method for efficiently factoring large semiprime numbers in a classical computational model has been discovered. 
In this paper, we demonstrate how a natural extension of the generalized approach to smoothness, combined with the separation of $2$-adic point orders, leads us to propose a factoring algorithm that finds (conjecturally) the prime decomposition $N = pq$ in subexponential time $L(\sqrt 2+o(1), \min(p,q))$. This approach motivated by the papers \cite{Len}, \cite{MMV} and \cite{PoZo} is based on a more careful investigation of pairs $(E,Q)$, where $Q$ is a point on an elliptic curve $E$ over $\Z _N$.
Specifically, in contrast to the familiar condition that the largest prime divisor $P^+(\ord Q_p)$ of the reduced order $\ord Q_p$ does not divide $\#E(\F_q)$ we focus on the relation between $P^+(\ord Q_r)$ and  the smallest prime number $l_{\min}(E,Q)$ separating the orders $\ord Q_p$ and $\ord Q_q$. We  focus on the ${\calE}_2$ family of even order elliptic curves over $\Z_N$ since then the condition $l_{\min}(E,Q)\le 2$ holds true for large fraction of points $(x,y)\in E(\Z_N)$. Moreover if we know the pair $(E,Q)$ such that $P^+(\ord Q_r)\le t<l_{\min}(E,Q)$ and $d=\max_{r\in \{p,q\}}(\ord Q_r)$ is large in comparison to $\min_{r\in \{p,q\}}|a_r(E)|\neq 0$ then we can decompose $N$ in deterministic time $t^{1+o(1)}$ by representing $N$ in base $d$.

\end{abstract}
\section{Introduction}

Nearly fifty years after RSA gained widespread recognition, it remains one of the most widely used public-key cryptographic schemes in the world. For instance, one of the well-known solutions for establishing a secret key between two parties without prior shared information is the handshake protocol used in SSL \cite{SSL} and TLS \cite{TLS} (which we will collectively refer to as TLS). 
Among the various schemes supported by these protocols, the handshake algorithm is specifically based on the RSA-PKCS-1v1.5 encryption scheme introduced in the PKCS \#1 v1.5 specification \cite{kaliski2016}. RSA can be utilized in two ways during the TLS handshake: as a key exchange method and as a signature scheme. As a key exchange method, RSA enables one party to encrypt a random value using the public key of the other party and transmit it securely.
On the other hand, as a signature scheme, RSA enables one party to sign a hash of the handshake messages using their private key and transmit it securely. Another critical aspect of TLS security is authentication, which involves verifying the identity of the involved parties. This authentication is typically achieved through certificates issued by trusted authorities, known as certificate authorities (CAs). These CAs vouch for the validity and authenticity of the certificates. RSA can be used to sign and verify these certificates, as well as to sign and verify handshake messages.
Another example is the SSH protocol, which is widely used for file transfer, communication in cloud services, and other computer-dependent tasks. SSH employs digital signature schemes to facilitate both server and client authentication. Alongside DSA and ECDSA, RSA DSS remains a widely adopted choice for securing this process (see \cite{bider2018rfc} for details).

Since its invention in 1976, RSA has been one of the most extensively studied cryptographic schemes, yet no efficient and feasible attack has been discovered so far. Consequently, it is not surprising that RSA has become the foundation for a new computationally hard problem called the RSA problem, which involves computing $[y^{1/e}\pmod{N}]$  for a uniformly chosen $y\in \ZZ_{N}^{*}$, an integer $e\in\ZZ_{>2}$ coprime to $\phi(N)$, and a modulus $N$ that is a product of two primes.This implies that RSA is secure as long as the RSA problem remains hard, and vice versa. It is worth noting that the RSA problem serves as the basis for the security of numerous cryptographic solutions, though a detailed discussion of this topic is beyond the scope of this paper. In fact, there exists a class of algorithms that exploit certain vulnerabilities in RSA parameters to break the RSA problem, and as a side effect, factor the associated modulus.
For example, Michael J. Wiener \cite{wiener1990cryptanalysis} observed that if the secret key satisfies $\sk<(1/3)\cdot N^{0.25}$, then there exists an efficient algorithm that solves both problems in polynomial time. Approximately 10 years later, Boneh and Durfee \cite{boneh1999cryptanalysis} improved Wiener's attack by demonstrating that the same result holds for $\sk< N^{0.293}$. 
The seminal work of Boneh and Durfee is one of the earliest heuristic extensions of Coppersmith’s methods \cite{coppersmith1996,coppersmith1996finding,coppersmith1997small} (see also Jutla \cite{jutla1998finding} and Bleichenbacher \cite{bleichenbacher1997security}). The idea behind this method is to construct a lattice related to RSA and then exploit the LLL algorithm to reduce the basis to an ordered short basis whose shortest vectors possess certain required properties. It is worth noting that Coppersmith's factoring method, presented in \cite{coppersmith1996finding}, is also used in this paper.

On the other hand, a hard computational problem that is commonly viewed as inherently tied to RSA is integer factorization. Obviously, the connection between the security of RSA and factoring is indisputable. However, it is important to bear in mind that the equivalence between factoring and the RSA problem does not hold. Specifically, access to an oracle solving the RSA problem does not imply the ability to solve the integer factorization problem. Consequently, factoring is much more general than the RSA problem and, in fact, influences (directly or indirectly) a much broader set of cryptographic schemes than just those based on the hardness of the RSA assumption. A good example is the DLP over finite fields, where an efficient factoring algorithm would significantly improve the index method. It turns out that, despite being one of the oldest mathematical problems, factoring cannot be solved efficiently by any known classical computing algorithm. In fact, the best algorithm for integer factorization depends on the size of the integer to be factored. For smaller integers, sieve methods and Pollard's rho algorithm \cite{pollard52monte,brent1980improved} are commonly used. For larger integers, the general number field sieve (GNFS) \cite{lenstra1990number} and the elliptic curve factorization (ECM) method \cite{Len} are more efficient. However, these algorithms have subexponential complexity. 
It should be noted that, although ECM is not the best-known algorithm in terms of complexity, it has opened up new perspectives for factoring arbitrary integers in time $L(\al_0, r)$, where $\al_0 =1/\sqrt 2$ and $p\mid N$ is the smallest prime dividing $N$. ECM, in particular, serves as the foundation for the considerations presented in this work. 

In this paper, we present a new approach to ECM factoring based on the concept of compositeness witnesses, adapted to Pollard's $p-1$ algorithm by B. Źrałek \cite{Zra} and further developed in \cite{PoRa}. Our approach builds on the notion of admissible elliptic curves and elliptic witnesses in the context of integer factorization, as previously explored in \cite{DryPom} and \cite{PoZo}. However, we refine this method by focusing on separating pairs $(E,Q)$ with distinct 2-adic orders $\operatorname{ord} Q_r$ for $r \in \{p,q\}$, as well as utilizing twisted elliptic curves $E^\tau$ to handle non-separating pairs $(E,Q)$. This leads to a more efficient computation of the orders of reduced points $\operatorname{ord} Q_r$ in terms of $t_{\min}(E,Q)$ whenever $(E,Q)$ is not a separating pair.

More specifically, we conduct an in-depth investigation of pairs $(E,Q)$ in the factorization of $N$ that are not $t$-separating, meaning that $M_t Q$ is not a finite point on $E$ over $\mathbb{Z}_N$ and that $l_{\min} > t$. This property can be verified in deterministic time $t(\log N)^{O(1)}$. We then focus on the corresponding set of $t$-consistent pairs $(E,Q)$ that satisfy condition (\ref{moduli}) below.

To summarize, we isolate the case $l_{\min}(E,Q) = 2$, as in this scenario, we always encounter a separating decomposition and can establish a precise characterization of the related pairs $(E,Q)$ in terms of Jacobi symbols. 
Rather than considering points $Q = (x,y)$ lying on the elliptic curve $E$ over $\mathbb{Z}_N$, as studied in \cite{MMV,DryPom2}, we find it more relevant to generate triples $(x,y,b_1) \in \mathbb{Z}_N^3$. We then state two theorems on separating and non-separating decompositions, respectively, and propose the algorithm $\mathcal{A}$ for factoring the RSA modulus $N$ under the Generalized Riemann Hypothesis (GRH) and a suitable conjecture regarding the distribution of $\#E(\mathbb{F}_r)$ values with large $B$-smooth divisors in the interval $I_r$, which lies within the Hasse interval $H_r$.

The algorithm applies to cases involving $B$-separating pairs $(E,Q)$, as well as those that are not $B$-separating but are $B$-consistent. Additionally, it applies to pairs $(E,Q)$ that are not $B$-consistent but satisfy the following condition:
\begin{equation}\label{verylargeord}
	\gcd\left(M_B(E,Q), E_r\right) \geq N^{1/2}/B,
\end{equation}
for some $r \in \{p,q\}$, where the multiplier $M_t(E,Q)$ is defined as the product of suitable powers of primes not exceeding $t$.

\section{Elliptic curves over $\Z_N$}

In this section, we recall basic facts about elliptic curves over $\Z_N$, where $N=\prod_{i=1}^s p_i$ is coprime to $6$
(see \cite{Len2,Len}).
The projective plane $\mathbb P^2(\Z_N)$ is defined as the set of equivalence classes of primitive triples in $\Z_N^3$ (i.e., triples $(x,y,z)$ with $\gcd(x,y,z,N)=1$) under the equivalence relation $(x_{1},y_{1},z_{1})\sim(x_{2},y_{2},z_{2})$ iff $(x_1, y_1,z_1) = u(x_2, y_2,z_2)$ for some unit $u\in \Z^*_N$. The equivalence class of $(x,y,z)$ is denoted by $(x:y:z)$ and is often simply referred to as a point. An elliptic curve over $\Z_N$ is given by the short Weierstrass equation $E: y^2z = x^3 + axz^2 +bz^3$, where $a,b\in \Z_N$ and
the discriminant $-16(4a^3 + 27b^2)$ is an element of $\Z_N^*$. The set of points satisfying this equation is denoted by \( E({\Z}_N) \), where, in particular, the points $(x:y:z)\in E(\Z_N)$, with $z\in \Z_N^*$, can be written as $(x/z: y/z:1)$ and are called finite points. Meanwhile, the point $O = (0:1:0)$, called the zero point, also belongs to $E(\Z_N)$.   
If $V(E(\Z_N))$ is the set of points $(x:y:z)$ in $E(\Z_N)$ such that  $z\not\in \Z_{N}^{*}$, then for each point $(x:y:z)\in E(\Z_N)\setminus V(E(\Z_N))$, the value $\gcd(z,N)$ is a nontrivial divisor of $N$.

Let $E(\F_{p_i})$ be the group of $\F_{p_i}$-rational points on 
the reduction $E\mod p_i$ for primes $p_i\mid N$. If $E(\Z_N)$ is the set of points in $\mathbb P^2(\Z_N)$ satisfying the equation of $E$, then by the Chinese Remainder Theorem, there exists a bijection 
\begin{align}\label{phi1} 
	\vp:  E(\Z_N) \to \prod_{i=1}^s E(\F_{p_i})
\end{align}
induced by the reductions mod $p_i$.  
The set $E(\Z_N)$ forms a group under addition, for which $\vp$ is a group isomorphism. In general, this addition can be defined using the so-called complete set of addition laws on $E$ 
(see \cite{Len2}).

To add two finite points $P,Q\in E(\Z_N)$, we can also use the same formulas as for elliptic curves over fields in the following case: 
for $\vp(P) = (P_1,\ldots,P_s)$ and  $\vp(Q) = (Q_1,\ldots,Q_s)\in \prod_i E(\F_{p_i})$, either $Q_i\neq \pm P_i$ for each $i$, or $Q_i =  P_i$ and $Q_i\neq   -P_i$ for each $i$.  
Then 
\begin{equation}\label{addel}
	\begin{cases}  x_{P+Q} = \lambda^2 - x_P - x_Q\\
		y_{P+Q} = \lambda(x_P - x_{P+Q}) - y_P,
	\end{cases}
\end{equation} 
where 
\begin{displaymath}
	\lambda = \begin{cases} \frac{y_Q- y_P}{x_Q-x_P} & \text{  if }  Q_i \neq \pm P_i \text{ for each } i \vspace{1mm}\\
		\frac{3x_P^2 +a}{2y_P} & \text{  if }  Q_i = P_i \text{ and } Q_i \neq -P_i \text{ for each } i. 
	\end{cases}
\end{displaymath} \vspace{1mm}
Let $P, Q\in E(\Z_N)$. If $R=P+Q$ is finite, then the formulas (\ref{addel}) give the coordinates of the resulting point $R$. Otherwise, either we find a nontrivial divisor of $N$, or we prove that all local orders $\ord R_i$ are equal for $i=1,2,..., s$ (see e.g. \cite{Len}, \cite{DryPom2} for details).

In what follows, we assume that $N=pq$ has two distinct prime divisors (both $>3$) and that $B=B(N)$ is fixed. We apply the above formulas to compute the point $mQ\in E(\Z_N)$. The computation of the finite point $mQ$ takes $O(\log m)$ addition operations in $E(\Z_N)$.
For $B$-smooth number $M=M_B$ represented as $m=p_k^{e_k}\ldots 3^{e_3}2^{e_2}$, where $e_i=e_i(m)$ is the highest exponent such that $p_i$ does not exceed $\min(p,q)+2\sqrt{\min(p,q)} +1$, the computation of $M_B Q$ requires
\begin{equation}\label {czeb}
	\ll\log N\sum_{i\le k}\log(p_i)=O\left(
	B\log N \right) 
\end{equation}
addition operations in $E(\Z_N)$ in view of the Prime Number Theorem. 

Let $N = pq$ be an RSA modulus, and assume that $p < q < \vartheta p$. Since the most difficult variant of the two-prime factorization problem arises when both $p$ and $q$ are close to $\sqrt{N}$, we accordingly restrict our consideration to primes that satisfy this condition. Consequently, for $r \in \{ p, q \}$, there exists a constant $c(\vartheta)$ such that $r = c(\vartheta)N^{1/2}$.

\subsection{Minimal decomposition $t$-multipliers for $(E,Q)$ }\label{sec::2.1}

Recall that the $l$-\textit{adic valuation} is defined as $\nu_l(m) = \max \{ k \in \mathbb{Z} \mid l^k \text{ divides } m \}$, that is, $ \nu_l(m) $ represents the exponent of $ l $ in the prime factorization of $ m $. We extend the concept of the $l$-adic valuation to the Hasse interval $H_{r}= \left\{ k \in \mathbb{Z} \mid \abs{k}\leq r+1+2\sqrt{r} \right\} $, where $r = c(\vartheta)N^{1/2}$. To this end, we introduce the set function $\nu_l$, expressed as
$\nu_l(H_r) = \max \{ k \in \mathbb{Z} \mid l^k \in H_r \cap \mathbb{Z}_{\geq 0} \}$,
while preserving the notation used for the $l$-adic valuation.
\begin{definition}
For $Q\in E(\ZZ_{N})$, $r=c(\vartheta)N^{1/2}$ and $t\in\ZZ_{>1}$, we define the number $M_t=M_t(E,Q,r)$ such that the following two conditions hold
\begin{enumerate}
	\item $M_{t}:=\prod_{l\in\mathcal{P}_{t}} l^{\nu_{l}(H_{r})}$, where $\mathcal{P}_{t}$ denotes the set of all primes less than or equal to $t$;
	\item $M_tQ$ is not a finite point of $E$ over $\Z_N$.
\end{enumerate}
Such an $M_t$ is referred to as the decomposition $N$ multiplier for the pair $(E,Q)$,or equivalently, as the $t$-multiplier for the pair $(E,Q)$. 
\end{definition}


By $l_{\min}(E,Q)$, we denote the smallest prime number $l$ for which the exponents $\nu_l(\ord Q_p)\neq\nu_l(\ord Q_q)$.  

\begin{definition}
By $t_{\min}(E,Q)$, we define the smallest $t$ for which $M_t$ is the decomposition $N$ multiplier for $(E,Q)$.
\end{definition}




\begin{definition}
	The pair $(E,Q)$ is called $B$-separating if both $t_{\min}(N,E,Q)=B$ and $l_{\min}(N,E,Q)\leq B$ hold.
\end{definition}

In the application, we consider randomly selected pairs $(E,Q)$ and choose the related (admissible) decomposition pairs $(E,Q)$ for $N$, with an emphasis on the associated separating and non-separating $t$-multipliers satisfying $t_{\min}(E,Q)\le B$.
In the non-separating case, it leads to the definition of $B$-consistent pairs $(E,Q)$ and the pairs $(E,Q)$ which do not satisfy the condition (\ref{moduli}) but satisfy the condition (\ref{verylargeord}).
 
\begin{definition}
	The pair $(E,Q)$ is called $B$-consistent if $t_{\min}(N,E,Q)=B$,  $l_{\min}(N,E,Q)> B$
   and the following condition holds:
	\begin{align}\label{moduli}
		d:=\ord{Q}_{p}=\ord{Q}_{q}\geq c(\vartheta)N^{{1}/{4}}\cdot\max\left( N^{{1}/{8}} , \, \min_{r\in\{p,q\}}\abs{a_{r}(E)}\right)\neq 0 ,
	\end{align}
	where $a_r(E)$ is the Frobenius trace of $E$ at $r\in\{p,q\}$.
	\end{definition}

	\begin{remark}
	It is easy to note that the condition (\ref{moduli}) implies the following 
	\begin{align}\label{moduli02}
		1\le\min_{r\in \{p,q\}}|a_r(E)|\le c(\vartheta)N^{-1/4}\max\left(\max_{r\in \{p,q\}}\ord Q_r, N^{{3}/{8}}\right).
	\end{align}
	\end{remark}

\section{Results on the separating decomposition}

In this section we focus on the separating decomposition of $n$ and prove the significant complexity bound for searching the separation pairs $(E,Q)$ for the family of  elliptic curves 
${\calE}_2$. The motivation to consider this family of elliptic curves comes from the \cite{MMV}, where the lower bound $cN(1+O(1/\sqrt{N}))$ for the number of points $(x,y)\in E(\Z_N)$ 
was proved with $c={41}/{225}$. The stronger bound with $c=27/80$ was obtained in \cite{DryPom2}.


\subsection{Result on searching the orders of reduced points in $E(\F_r)$}

In the paper \cite{DryPom}, a deterministic algorithm for finding the decomposition $N=pq$, given the $t$-multiplier for the pair $(E,Q)$ (i.e. the decomposition $N$ multiplier) with $l_{\min}(E,Q)\le t$, was presented with a deterministic runtime of $t^{2+o(1)}$.
In the next section, we improve this result to a deterministic runtime of $t^{1+o(1)}$.
This implies fast computation (in terms of $t$) of the orders of reduced points, given the $t$-multiplier for the pair $(E,Q)$, provided that $l_{\min}(E,Q)\le t$. Let $E$ be an elliptic curve over $\Z_n$. The result is based on the following lemma:
\begin{lemma}\label{binary}
	Assume that we are given $t$-multiplier $M_t$ for the pair $(E,Q)$ for some $t\le B$.
	Then, in deterministic time $B$, one can compute the decomposition $N=pq$ or find both reduced orders $\ord Q_p$ and $\ord Q_q$, which must be equal.
\end{lemma}
\begin{proof}
	Let $r\mid N$ and $M_t$ be the multiplier for $(E,Q)$, where $t\le B$.
	Assume that $\ord Q_r= 2^{\nu_2}\cdots p_u^{\nu_u}$, where $p_u=P^+_{\min}:=\min_{r\in \{p,q\}}P^+(\ord Q_r)\le B$. In deterministic time $B^{1+o(1)}$, we find the smallest $t$ such that $M_tQ$ is not a finite point in $E$ over $\Z_N$.
	It is clear that $t=p_u:=P^+_{\min}$. Let $P^+_{\min}:=P_0$  $p_{u-1}:=P_1, ~p_{u-2}=P_2$, and so on.
	
	We will show how to find $P_1\in [2,P_0]:=I_1$ (if exists) in the related exponent $\nu_1$ , which divides exactly the related order $\ord Q_r$, in deterministic time $B^{1+o(1)}$. Following an analogous procedure for the intervals $[2,P_1)$, then $[2,P_2)$, and so on, the algorithm will terminate in deterministic time $\omega(E_r)B^{1+o(1)}=B^{1+o(1)}$, since $E_r\le 2r\ll\log N/\log\log N$.
	The computation resembles a binary search. We know that $I_1=I_1^-\cup I_1^+$, split by the median point $t_1$ of $I_1$. If $P_2$ lies in $I^+$, then the point $M_{t_1}P_1^{\nu_1}Q\in E(\Z_n)$ is finite; otherwise it is not. Thus, we can determine the correct case in deterministic time 
	$$
	(\log N)^{O(1)} \log \left(M_{t_1}P_1\right)\ll (\log N)^{O(1)} \left(t_1 B^{\nu_1}\right)\ll (\log N)^{O(1)} \left(t_1 \sqrt n\right)
	\ll B^{1+o(1)},
	$$ 
	provided $B\gg (\log N)^{O(1)}$. Then, we find the appropriate exponent such that $P_2^{\nu_2}$ divides the suitable order $\ord P_r$, checking all possible cases where $\nu_2\le \log_2 N$.  The correct values of $P_2$ and $\nu_2\le \log_2 N$ will be determined after at most $O(\log^2 N)$ trials, since the length of the subintervals $I_i$ (for $i\le \log _2 N$) tends to $1$. This completes the proof 	by continuing the search of $P_i^{\nu_i}$ for $i\le \log_2 N$.
\end{proof}

\subsection{$2$-adic separation for the family ${\calE}_2$}

From now on, we assume that $N$ is sufficiently large positive integer coprime to $6$.
In what follows, we  consider the elliptic curves over $\Z_N$ given by either the Weierstrass equation
$y^2=f_{\bar{B}}(x)$ or $y^2=f_{\bar{b}}(x)$, where $\bar{B}$ is either an element of $(B_1,B_2)\in \Z_n^2$ or $\bar{b}=(b_1,b_2)$, with $(b_1,b_2)\in \Z_n^2$ and
either
\begin{align*}
f_{\bar{B}}(x)=x^3+B_1 x+B_2,
\end{align*}
or
\begin{align*}
f_{\bar{b}}(x)=(x-b_1)(x-b_2)(x+b_1+b_2).
\end{align*}

It is clear that the elliptic curve $E_{\bar{b}}$ over $\Z_N$ has the short Weierstrass form $E_{\bar{B}}$ over $\Z_N$ with suitable $B_1=B_1(b_1,b_2)$ and $B_2=B_2(b_1,b_2)$.

\begin{lemma}\label{knapp}(see Theorem~4.2, \cite{knapp1992elliptic}) \label{kn} 
Let $E$ be an elliptic curve over the field $\F_r$ of characteristic $\neq 2,3$, 
given by the equation 
\begin{align*}
y^2 =(x-b_1)(x-b_2)(x+b_1+b_2) 
\end{align*}
with $b_1,b_2\in\F_r$. For $(x_2,y_2)\in E(\F_q)$, there exists $(x_1, y_1) \in 
E(\F_r)$ with $2(x_1,y_1)= (x_2,y_2)$ if and only if both $x_2-b_1, x_2-b_2$ and $x_2+b_1+b_2$ are squares in $\F_r$. 
\end{lemma}

\begin{theorem} \label{jacobi}
Let $N=pq$, $E$ is elliptic curve over $\Z_N$ and $\nu_2(E_p)\ge \nu_2(E_q)$. Furthermore, assume
that 
\begin{align}\label{e1}
\left( \frac{x-b_1}{N} \right)=-1,
\end{align}
and that $b_2 \pmod{N}$ satisfies the quadratic equation
\begin{align}\label{e2}
\frac{y^2}{x-b_1}=(x-b_2)(x+b_1+b_2) \pmod{N}.
\end{align}
Moreover, if
\begin{align}\label{e0}
  \gcd\left((b_2-b_1)(2b_1+b_2)(2b_2+b_1), N\right)=1,
\end{align}
and
\begin{align}\label{e3}
\left( \frac{x-b_1}{p} \right)=-1, ~~~
\left( \frac{x-b_1}{q} \right)=1.  
\end{align}
Then we have
\begin{align}
\nu_2(\ord Q_p )> \nu_2(\ord Q_q),
\end{align}
if and only if
\begin{equation}\label{e4}
\left(\frac{x-b_2}{q}\right)=1.
\end{equation}

\end{theorem}

\begin{proof}
Assume that $\bar{b}$ satisfies the equation $y^2=f_{\bar{b}}(x,y) \pmod{N}$, that is, $(x,y)\in E_{\bar{b}}(\Z_N)$, and $b_1$ satisfies the equation (\ref{e1}). Assume further that $b_2$ satisfies the equation (\ref{e2}). Then, by Lemma \ref{knapp}, we conclude that the equation
$2(x_0,y_0)=(x,y)$
has a solution in $\F_r^2$ if and only if at least one of the numbers
$(x-b_1), (x-b_2), (x+b_1+b_2)$ is a square in $\F_r$.
Hence, $(x,y)$ has maximal $2$-adic order in $\F_p$, but does not when $x-b_2$ is a square modulo $p$ (by the condition on $b_2$, it follows that  $x+b_1+b_2$ must also be a square in this case). 
Now, using the assumption that $\nu_2(E_p)\ge \nu_2(E_q)$, we arrive at the desired conclusion.
\end{proof}
The results of the above theorem justify the following definition.

\begin{definition}
We define a triple $T=(x,y,b_1)\in \Z_N^3$ as admissible if it satisfies the following conditions:
\begin{itemize}
\item[(i)] The coefficient $b_1$ satisfies the equality (\ref{e1}) for all $x,y,b_1\in \Z_N$.
\item[(ii)] The element $b_2$ modulo $N$ satisfies the equation (\ref{e2}) for given values of $x,y$, and $b_1$ modulo $N$.
\item[(iii)] The condition (\ref{e0}) holds with $b_2=u$ as a solution of equation (\ref{ee}) modulo $N$.
\end{itemize} 
\end{definition}

\begin{remark}
It should be noted that $b_{2}=b_{2}(b_{1})$. Thus, due to this, instead of $E_{(b_{1},b_{2})}(\Z_{N})$ we will write $E_{b_{1}}$. 
\end{remark}


\begin{theorem}
Let $N=pq$ and assume that $\nu_2(E_p)\ge \nu_2(E_q)$. 
Then, the probability that a randomly chosen triple $T=(x,y,b_1)\in\Z_{N}^{3}$ is admissible and additionally satisfies the condition that the point $Q=(x,y)\in E_{\bar{b}}(\Z_N)$ has $\nu_2(\ord Q_p)\neq \nu_2(\ord Q_q)$, is greater than or equal to $N^{-3}\cdot\sum_{T\in S}1$.
Here, $S$ consists of all admissible triples that satisfy (\ref{e3}) and for which 
$\Delta=\Delta(x,y,b_1):=(2x-b_1)^2- 4\left(x^2+xb_1-y^2(x-b_1)^{-1} \right)$ is a quadratic residue modulo $N$. Additionally, a solution $u$ to the quadratic equation
\begin{equation}\label{ee}
f(u):=u^2+b_1u -\left( x^2+xb_1-y^2(x-b_1)^{-1} \right)=0\mod q
\end{equation}
must satisfy $\left( \frac{u}{q}\right)=1$. 
Furthermore, the condition (\ref{e0}) holds for $b_2=u\mod N$.
\end{theorem}
\begin{proof}
The existence of a solution to equation (\ref{ee}) follows from the fact that  $\Delta$ is a square modulo $N$.
Moreover, at least one of the roots $u= \frac{1}{2}(-b_1\pm \sqrt \Delta)$ of (\ref{ee}) satisfies the equation $\left(\frac{u}{q}\right)=1$.
Furthermore, by assumption, the condition (\ref{e0}) holds. This completes the proof of the theorem. 
\end{proof}

\begin{lemma} (Coppersmith) \label{Cop} (see \cite{coppersmith1996finding})
If we know $N$ and the highest-order $(1/4)(\log_2 N)+O_\vartheta(1)$ bits of $q$, where $p<q<\vartheta p$, then we can determine $p$ and $q$ in polynomial time in $\log N$.
\end{lemma}
As a corollary, we deduce:
\begin{corollary}
Assume that $M_B$ is the multiplier for $(E,Q)$ and assume that $\ord Q_r\ge E_r^\beta$
for some $\beta>1-2\frac{\log B}{\log N}$. Then, the  decomposition of $N$ can be computed in deterministic time $B (\log N)^{O(1)}$.
\end{corollary}

\begin{proof}
If $|E|_r$ is known, then $\frac{1}{2}\log r=\frac{1}{4}\log_2 N+O_\vartheta (1)$ is also known, allowing us to apply the aforementioned lemma to reach the conclusion. We now have
$$
\frac{|E|_r}{\ord Q_r}\le E_r^{1-\beta}<2N^{\frac{1-\beta}{2}}
$$
for some $r\in \{p,q\}$. Using the inequality for $\beta$, the above expression is $O(B)$, and thus the conclusion follows.
\end{proof}

We recall that in factoring with the oracle \textsf{OrdEll}, the result proved in \cite{DieJim} states that given the decomposition of $E_N$ one can find $p$ and $q$ in randomized polynomial time with probability $1-\ve$ for arbitrary fixed $\ve>0$.

\section{Results on non-separating decomposition of $N$}
In this section, we prove the related results on the non-separating decomposition 
by proving that if 
$t_{\min}(E,Q):=B$ is not a separating multiplier, but is
$B$-consistent, then $N = pq$ can be decomposed in the {similar} time as before, i.e., $B(\log N)^{O(1)}$ {under the assumption of Generalized Riemann Hypothesis (GRH)}.

\begin{lemma}\label{infty}
Assume that $B=t_{\min} (E,Q)$.
Then in deterministic time $B (\log N)^{O(1)}$ we either find a nontrivial divisor of $N$, or {compute}
\begin{equation}\label{d}
d:=s_B(\ord Q_p)=\ord Q_p=s_B(\ord Q_q)=\ord Q_q.
\end{equation}
{Here}, $s_B(m)$ denotes the largest $B$-smooth factor of $m$.
\end{lemma}

\begin{proof}
Assume that $B = t_{\min}(E, Q)$. Referring to Section $2$, we compute $M_B Q$ where $M_B = 2^{\nu_2} \cdots p_u^{\nu_u} Q$. This computation yields two possibilities: the greatest common divisor (GCD) is either a nontrivial divisor of $N$, or the GCD is equal to $N$. In the latter case, the pair $(E, Q)$ does not separate, hence the conclusion follows. This requirement is supported by Lemma \ref{binary}.
\end{proof}

\begin{definition}
Let $E$ be an elliptic curve over $\Z_N$ given by the short Weierstrass equation
$$
y^2=x^3+b_1x+b_2.
$$
where $4b_1^3+27b_2^2\neq 0 \mod n$ and $\tau\neq 0\mod n$.
Then the twisted elliptic curve $E^\tau$ over $\Z_n$ is defined by the equation
$$
y^2=x^3+\tau^2b_1x+\tau^3 b_2.
$$
\end{definition}

\begin{lemma} \label{znak}(see Theorem 1.4 in \cite{Lamzouri2015})
	Assume GRH. Let $E$ be an elliptic curve $E$ over $\Z_N$, and let $(\mu_1,\mu_2)$ be an arbitrary pair in $ \{(\pm1,\pm1)\}$. Then  there exists $\tau \ll \log^2 N$ such that for the twisted elliptic curve $E^\tau$, we have $(a_p(E^\tau), a_q(E^\tau))=(\mu_1a_p(E),\mu_2a_q(E))$, where $a_{p}(E)$ and $a_{q}(E)$ are the Frobenius traces of $E$ at primes $p$ and $q$, respectively.
\end{lemma}

%


%
\begin{theorem}\label{consistent}
	Assume GRH. {Let $N$ be an 
    RSA modulus with sufficiently large prime factors $p,q=O(\sqrt N)$.
    Let $(E,Q)$, $B=t_{\min}(E,Q)$, and $d$ be as specified in Lemma \ref{infty}, and assume that $l_{\min}(E,Q)>B$.
	Then, there exists a deterministic algorithm $\calA$ that decomposes $N=pq$, in deterministic polynomial time $(\log N)^{O(1)}$, provided that the  pair $(E,Q)$ is $B$-consistent.}
\end{theorem}

\begin{proof}
	We will prove that one can decompose $N=pq$ in deterministic time $B(\log N)^{O(1)}$ where $B=t_{\min}(E,Q)$, either if $l_{\min}(E,Q)\le B$ or if $l_{\min}(E,Q)>B$ and the pair $(E,Q)$ is $B-$consistent. The first case is straightforward since the pair  $(E,Q)$ is then a $B-$separating pair. Assume that $(E,Q)$ is not a $B$-separating pair, and let $d$ be specified in Lemma \ref{infty}. We will demonstrate that the decomposition $N=pq$ can be computed in deterministic polynomial time $(\log N)^{O(1)}$ if the condition (\ref{moduli}) holds true. 
	
	 In view of Lemma \ref{znak} for given $E$ over $ \Z_N$,  we will determine in deterministic time $O(\log^2N)$ the twisted parameter $\tau$ such that
	\begin{align}
		d:=\ord{Q}_{p}=\ord{Q}_{q}\geq c(\vt)
		N^{{1}/{4}}\cdot\max\left( N^{{1}/{8}} , \, \max_{r\in\{p,q\}}a_{r}(E^\tau)\right)\neq 0 ,
	\end{align}
	where  $c(\vt)=\max(2c'(\vt),c''(\vt))$ (see below) depends only on $\vt$.
	Hence $d\ge c(\vt)N^{3/8}$ and $1\le \min_{r\in \{p,q\}}a_r(E^\tau)\le \frac{d}{c(\vt)}N^{-1/4}$, placing us in a position to apply a similar argument as in \cite{PoZo}. Specifically, by representing $N$ in base $d$, we have
	\begin{align*}
		N=c_0+c_1d+c_2d^2,
	\end{align*}
	where $c_i \in [0,d-1]$ and $c_2\neq 0$. 
	Moreover, since $d\mid \gcd(E^\tau_p,E^\tau_q)$, we can write
	\begin{align*}
		E^\tau_p=p+1-a_p(E^\tau)=dr_p, \, E^\tau_q=q+1-a_q(E^\tau)=dr_q,
	\end{align*}
	and letting $t_r:=a_r-1\ge 1$ for $r\in \{p,q\}$, it follows that
	\begin{equation}\label{n=pq}
		N=pq=(dr_p+t_p)(dr_q+t_q)=c_0+c_1d+c_2d^2.
	\end{equation}
	If the representation of $N$ in base $d$ is unique (i.e., the coefficients $c_i\in [0,d-1])$), then we can recover (see below) the values  $r_p, r_q, t_p, t_q$ and consequently $p$ and $q$, in deterministic $\text{polylog}(N)$ time. 
	
	 The condition $0\le r_pr_q\le d-1$ follows from the inequality $(p-2\sqrt p)(q-2\sqrt q)\le E_pE_q\le d^2(d-1)<d^3$, along with the fact that  $d\ge c(\vt)N^{3/8}> N^{1/3}$.
	The remaining two conditions follow from the $B$-consistency of the pair $(E,Q)$.
	Namely, by $B$-consistency of a pair $(E,Q)$ we have
	$$
	1\le t_pt_q\le {c'(\vartheta)}(d/N^{1/4})N^{1/4} \le d,
	$$
as required. Moreover we have that
	$$
	r_pt_q+r_qt_p\le \max(t_p,t_q)\left( \frac{E_p+E_q}{d}\right)
	\le c''(\vt)N^{1/4}(N^{1/2}/d) \le c''(\vt)N^{3/4}/d\le d,
	$$
	since $d \geq c(\vt)N^{3/8}$.  Therefore, the polynomial $f(x)=c_0+c_1x +c_2x^2=(xr_p+t_p)(xr_q+t_q)$ has roots  
		$x_1=-\frac{t_p}{r_p}=-\frac{k_1t_p^*}{k_1r_p^*}$ 
		and $x_2=-\frac{t_q}{r_q}=-\frac{k_2t_q^*}{k_2r_q^*}$, where $k_i$ are the greatest common divisors of $t_p/r_p$ and $t_q/r_q$ for $i=1,2$ respectively. 
		
		 We will prove that $k_i=1$ for $i=1,2$. Suppose, on the contrary, that $2\le k_1\mid dr_p+t_p=p$ or $2\mid dr_q+t_q=q$ in view of (\ref{n=pq}). This would imply that $pq$ must be even which, contradicts the assumption that $N$ is odd. Hence, we conclude that $x_1=-\frac{t_p}{r_p}$ and $x_2=-\frac{t_q}{r_q}$ are irreducible. Therefore, we can recover the values of $r_p, \, r_q$ from the denominators of $x_1$ and $x_2$ and $t_p, \, t_q$ from their numerators. Finally, we we compute $p=dr_p+t_p$ and $q=dr_q+t_q$, as required. 
		
	\end{proof}

\section{Algorithm $\calA$ and possible acceleration of decomposition $N=pq$}

Here we propose an algorithm factoring the RSA numbers and make the brief analysis computational savings in case when $(E,Q)$ is not separating pair for $N$.

\subsection {Description of algorithm $\calA$}

It is clear that in applications, $B$ cannot be too large, as it is the minimal non-separating multiplier for $(E,Q)$, nor can it be too small, since the maximal $B$-smooth divisor of $\ord Q_r$ must be large. We begin with the following definition:

\begin{definition}
	The pair $(E,Q)$ is called $B$-decomposition pair for $N$ if at least one of the following conditions holds:
	\begin{itemize}
		\item[(i)] 
		$\max\left(t_{\min}(E,Q,N), \,l_{\min}(E,Q,N)\right)\le B$;
		\item[(ii)] 
		$t_{\min}(E,Q,N)\le B<l_{\min}(E,Q,N)$ and $(E,Q)$ is $B$-consistent pair;
		\item[(iii)]
		$\gcd(M_B(E,Q), E_r)\ge \abs{E}_r/B$ for some $r\in \{p,q\}$.
	\end{itemize}
\end{definition}
\begin{remark}
	It must be emphasized that (ii) implies $\ord{Q_{p}}=\ord{Q_{q}}$ and, furthermore, it is equivalent to $d\geq c(\vartheta)\cdot n^{1/2}/{B}$, where $c(\vartheta)$ is a constant.
\end{remark}
The knowledge of $B$-decomposition pair $(E,Q)$ implies that one can determine the factorization $N=pq$ in deterministic time $(\log N)^{O(1)} B$. 
The choice of an optimal $B$ should balance the cost of computing the minimal $B-$multiplier for $N$ and the probability of finding the $B$-decomposition pair $(E,Q)$ for $N$.

\vspace{0.3cm}
\noindent {{\bf{Algorithm $\calA$}}}\\
The randomized algorithm $\calA$ for decomposing $N$ performs the following steps:
\noindent \textbf{Input}: The number $N$ to be factored.\\
\noindent \textbf{Output}: Decomposition $N=pq$.
\begin{enumerate}
	\item Randomly choose a triple $T=(x,y,b_1)\in \Z_N^3$ until it is an admissible triple. 
    \item Select an admissible pair $(E,Q)$ and all the twists $E^{\tau}$, where $\tau=O(\log^{2}{N})$ is generated as described in \cite{Lamzouri2015}.
	\item Compute $t_{\min}(E,Q):=B$ as presented in Section (\ref{sec::2.1}).
	\item Return the nontrivial divisor $d\mid N$ if $t_{\min}(E,Q)$ is a separating multiplier; otherwise, proceed to Step 4.
	\item Compute the reduced orders $\ord Q_p$, $\ord Q_q$ satisfying the conditions: $$d:=\ord Q_p = \ord Q_q.$$
	\item If $d\le N^{3/8}$, (i.e. consistency does not occur), then return to step 1; otherwise, go to step 6.
	\item Represent $n=c_0+c_1d+c_2d^2$ and perform the computations described in Theorem \ref{jacobi} to search the decomposition $N=pq$. If this fails, go to step 7.
	\item If $d>\sqrt N/B$, then for all $k\le B$ compute $k d$ and apply the Coppersmith method to search for the decomposition of $N$ as $N=pq$. If it fails, return to step 1.
\end{enumerate}

\subsection{Assumed conjecture and analysis of algorithm complexity}

Assume that the triples $T=(x,y,b_1)$ are generated with uniform distribution and that the orders of curves $E_{b_1}(\F_r)$ are almost uniformly distributed in the interval $I_r:=(r+1-\sqrt r, r+1+\sqrt r)\subset H_{r}$
(see \cite{Len,clozel2008sato} and Theorem 5.2 in \cite{konyagin1997primes}). 
More precisely let $\al>0$,  $B=L(\al,x):=\exp(\al\sqrt{\log x\log\log x} )$. We assume the following

\smallskip
\noindent{\bf{Conjecture.}}
Let 
$$
v=v(x,B,\beta):=\#\{ m\in \Z: |m-(x+1)|<\sqrt x \,\, \mbox{and} \,\, s_B(m)\ge m^\beta\}\ge 3 
$$
and $f(x,B,\beta)=\frac{v(x,B,\beta)}{(2[\sqrt x]+1)}$ denotes the probability that a random integer in $I_x$ is $(B,\beta)$-smooth number.
Then we conjecture that 
$$
f(r,L(\al,r),\beta) \ge L^{-1}\left(\frac{1-\theta(1-\beta)}{2\al},r\right),
$$
for some $\theta\ge 0$, as prime $r$ tends to infinity. We indicate that this time $\theta $ does not depend on $\beta$.

For our purposes, we consider two cases $\beta=1$ and $\beta=3/4$. The first case applies when we count the elliptic curves $E_{b_1}$ of $B$-smooth order over $\F_r$, while the second  applies when $|E_{b_1}|_r$ is a $(B,3/4)$-smooth number for each $r\in \{p,q\}$. In the first case, we perform the standard evaluation, searching for the separating pair $(E,Q)$ for $N$ with computational costs $L(\frac{1}{2\al}+o(1),r)$ (the algorithm terminates at step 3). In the second case, we aim to discover the $B$-consistent pair $(E,Q)$ with a probability not exceeding $f(r,B,3/4)^2$. This conclusion is drawn from the fact that if $(E,Q)$ is not a separating pair, then both reduced orders $\ord Q_r$ are greater than or equal to $ N^{3/8}$. Consequently, the related group orders $|E|_r$ must possess the same $B$-smooth factor, not less then $N^{3/8}$, yielding the {the upper bound $f(r,B,3/4)^2$ for the related probability} . We encounter this  situation in step 6 of the algorithm $\calA$ with additional polynomial-time computational overhead. Finally, in step 7, the additional computational overhead is equivalent to $B^{1+o(1)}$ as $r$ tends to infinity. 

\begin{example}
We aim to factorize the number 
$N = 3839985129719.$ To achieve this, we define $E=E_{\bar{b}}$, where $\bar{b} = \left(1594604, 450302\right)$ and $Q = \left(540525859015,\right.$ $\left. 1621377667969\right)$.  
It is straightforward to verify that the pair  $(E,Q)$ is $B$-consistent for $B=3$. Next, we compute $M_{t=B}:=\prod_{l\in\mathcal{P}_{B}} l^{\nu_{l}(H_{r})}=557256278016$. Determining the minimal multiplier, we find that $d=\ord Q_p = \ord Q_{q} = 2^{7}3^{7}$. Expressing $N$ in base $d$, we obtain $N = 49d^{2} + 504d + 1271$. Now solving the  system of equations
\begin{align*}
\begin{cases}
t_{p}t_{q} &= 1271, \\
r_pt_{q} + r_qt_{p} &= 504, \\
r_{p}r_{q} &= 49, \\ 
N &= (279936r_p + t_p)(279936r_q+t_q). 
\end{cases}
\end{align*}
We compute $r_p = r_q = 7, t_p = 31, t_q =41$ and recover $p=1959583=dr_p+t_p$ and $q=dr_q+t_q=1959593$. Consequently, we obtain the factorization $N=1959583 \cdot 1959593$.

\end{example}

\section{Conclusions}
We examine the relationship between the largest prime divisor $P^+(\ord Q_r)$ of the reduced order $\ord Q_r$ and the smallest prime $l_{\min}(E, Q)$ that separates $\ord Q_p$ and $\ord Q_q$. Unlike the usual condition that $P^+(\ord Q_p)$ does not divide $\#E(\mathbb{F}_q)$, our focus is on the $\mathcal{E}_2$ family of even-order elliptic curves over $\mathbb{Z}_N$, where $l_{\min}(E, Q) \leq 2$ holds for a large fraction of points $(x, y) \in E(\mathbb{Z}_N)$. Furthermore, given a pair $(E, Q)$ such that $P^+(\ord Q_r) \leq t < l_{\min}(E, Q)$ and $d = \max_{r \in \{p, q\}}(\ord Q_r)$ is large relative to $\min_{r \in \{p, q\}} |a_r(E)| \neq 0$, we can deterministically factor $N$ in time $t^{1+o(1)}$ by expressing $N$ in base $d$.

\vspace{0.5cm}
\noindent \textbf{Acknowledgments.} The authors would like to thank Olgierd Żołnierczyk, a PhD student, for his contributions to Example 1.

%
%
%
 \bibliographystyle{splncs04}
 \bibliography{mybibliography}

\end{document}